\documentclass[%
 aip,
 amsmath,amssymb,
 reprint,%
]{revtex4-1}

\usepackage{graphicx}
\usepackage{dcolumn}
\usepackage{bm}
\usepackage{booktabs}
\usepackage{xcolor}
\usepackage[utf8]{inputenc}
\usepackage[T1]{fontenc}
\usepackage{mathptmx}
\usepackage{etoolbox}
\usepackage{newunicodechar}
\usepackage{physics}
\newunicodechar{⁺}{\textsuperscript{+}}
\usepackage{booktabs} 
\usepackage[margin=1in]{geometry}

\makeatletter
\def\@email#1#2{%
 \endgroup
 \patchcmd{\titleblock@produce}
  {\frontmatter@RRAPformat}
  {\frontmatter@RRAPformat{\produce@RRAP{*#1\href{mailto:#2}{#2}}}\frontmatter@RRAPformat}
  {}{}
}%
\makeatother
\begin{document}

\preprint{AIP/123-QED}

\title{Physically-Constrained Autoencoder-Assisted Bayesian Optimization for Refinement of High-Dimensional Defect-Sensitive Single Crystalline Structure}

\author{Joseph Oche Agada}
    \email{joe88data1@gmail.com.}
\affiliation{Bredesen Center for Interdisciplinary Research, University of Tennessee, Knoxville, USA, 37996}
\affiliation{Center for Advanced Material Science and Manufacturing, University of Tennessee, Knoxville, TN 37996, USA}%
\author{Andrew McAninch}%
\affiliation{Center for Advanced Material Science and Manufacturing, University of Tennessee, Knoxville, TN 37996, USA}
\affiliation{Department of Physics and Astronomy, University of Tennessee, Knoxville, TN 37996, USA}
\author{Haley Day}
\affiliation{Georgia Institute of Technology, Atlanta, GA 30332, USA}
\author{Yasemin Tanyu}
\affiliation{Virginia Polytechnic Institute and State University, Blacksburg, VA 24061, USA}
\author{Ewan McCombs}
\affiliation{Department of Materials Science and Engineering, University of Tennessee, Knoxville, TN 37996, USA}
\author{Seyed M. Koohpayeh}
\affiliation{Department of Physics and Astronomy, Institute for Quantum Matter, Johns Hopkins University, Baltimore, MD 21218}
\affiliation{Department of Materials Science and Engineering, Johns Hopkins University, Baltimore, MD 21218}
\affiliation{The Ralph O’Connor Sustainable Energy Institute, Johns Hopkins University, Baltimore, MD 21218}
\author{Brian H. Toby}
\affiliation{Advanced Photon Source, Argonne National Lab, Lemont, IL 60439, USA}
\author{Yishu Wang}
    \email{wangyishu@utk.edu.}
\affiliation{Center for Advanced Material Science and Manufacturing, University of Tennessee, Knoxville, TN 37996, USA}
\affiliation{Department of Physics and Astronomy, University of Tennessee, Knoxville, TN 37996, USA}
\affiliation{Department of Materials Science and Engineering, University of Tennessee, Knoxville, TN 37996, USA}
\author{Arpan Biswas}
    \email{abiswas5@utk.edu.}
\affiliation{Center for Advanced Material Science and Manufacturing, University of Tennessee, Knoxville, TN 37996, USA}
\affiliation{University of Tennessee-Oak Ridge Innovation Institute, Knoxville, TN 37996, USA}

\date{\today}

\begin{abstract}
Physical properties and functionalities of materials are dictated by global crystal structures as well as local defects. To establish a structure-property relationship, not only the crystallographic symmetry but also quantitative knowledge about defects are required. Here we present a hybrid Machine Learning (ML) framework that integrates a physically-constrained variational autoencoder (pc-VAE) with different Bayesian Optimization (BO) methods to systematically accelerate and improve crystal structure refinement with resolution of defects. We chose the pyrochlore structured Ho$_2$Ti$_2$O$_7$ as a model system and employed the GSAS-II package for benchmarking crystallographic parameters from Rietveld refinement and for training data generation. However, the function space of these material systems is highly non-linear, which limits  optimizers, such as in traditional Rietveld refinement, into trapping fits at local minima. Also, these naive methods do not provide an extensive learning about the overall function space, which is essential for large space, large time consuming explorations to identify various potential regions of interest. Thus, we present the approach of exploring the high-Dimensional structure parameters of defect-sensitive systems via pretrained pc-VAE assisted Bayesian optimization and Sparse Axis Aligned Bayesian Optimization. The pc-VAE, designed and trained on physically plausible Ho$_2$Ti$_2$O$_7$ structure models, projects high-Dimensional diffraction data consisting of thousands of independently measured diffraction orders into a low-D latent space while enforcing scaling invariance and physical relevance of the latent space. In this proposed design of closed-loop autonomous exploration, we aim to minimize the $\chi^2$ errors, also known as L2 norm, in the real and latent spaces separately between experimental and simulated diffraction patterns, thereby steering the refinement towards potential optimum in the parameter space of crystal structures. We investigated and compared the results among different methods such as pc-VAE assisted BO, non pc-VAE assisted BO, and Rietveld refinement. The result shows that the methodology can be generalized to other complex materials where ultra-precise determination of structural defects is needed to reveal subtle structure–property relationships, highlighting a new paradigm for integrating crystallography with machine learning to accelerate discoveries and characterizations of magnetic materials. %
 
\end{abstract}

\maketitle

%

\section{\label{sec:level1}INTRODUCTION}

Artificial intelligence (AI) and machine learning (ML) have emerged as transformative tools in the automation of materials characterization, addressing long-standing challenges associated with expert-driven and time-intensive analysis of diffraction, microscopy, and magnetic data. Early applications relied on classical machine-learning models such as support vector machines and decision trees, which used engineered diffraction features to classify crystal structures. While these methods demonstrated interpretability and reasonable accuracy, they were fundamentally constrained by the need for manual feature extraction \cite{seko2020}. A major breakthrough occurred with the adoption of deep learning, especially convolutional neural networks (CNNs), which enabled end-to-end learning from raw X-ray diffraction (XRD) patterns. Landmark studies demonstrated that CNNs could classify crystal systems with accuracies approaching 95\% \cite{park2017}, outperforming classical approaches and establishing deep-learning as the dominant method for diffraction analysis. Subsequent work expanded these models to handle multi-phase mixtures, noisy experimental data, and property prediction directly from diffraction patterns \cite{lee2020, lee2022}. Transfer learning and physics-informed data augmentation further improved robustness and generalization to experimental conditions \cite{oviedo2019, lee2023}. Beyond CNNs, emerging methods such as vision transformers provide global attention mechanisms that enhance interpretability and long-range pattern recognition in spectral and diffraction datasets \cite{chen2024b}. Moving forward, active learning method such as Bayesian optimization frameworks have been applied to explore computationally expensive material spaces, to attain convergence in minimal iterations \cite{Biswas2021a, Morozovska2021, Morozovska2022, Kalinin2020BOGenerative}.

ML is particularly advantageous for studying materials in which complex interactions give rise to emergent behavior. Quantum magnetic systems, where competing exchange interactions, known as magnetic frustration, foster unconventional ground states and exotic excitations with potential relevance to quantum information technologies~\cite{Balents2010, Moessner2006}, represent a prominent example. Due to the intrinsic complexity of these interactions and phenomena, an increasing amount of effort has been made to leverage ML approaches such as generative models to accelerate their discovery and understanding. For example, diffusion models and transformer-based systems such as DiffractGPT can propose crystal structures consistent with observed diffraction patterns \cite{guo2024b, choudhary2025}. Another recent example, Samarakoon et al. employed an unsupervised variational autoencoder model to analyze neutron-scattering data in order to enable automated identification of correlations in Dy$_2$Ti$_2$O$_7$ pyrochlore and provide new thermodynamic insights \cite{Samarakoon2020}. Furthermore, they applied the Gaussian process regression (GPR) to infer the optimal parameters of the Hamiltonian model by comparing simulated and experimental scattering data. Kwon et al. applied deep convolutional neural networks (CNNs) to efficiently search for the ground-state configurations of complex spin-ice systems, successfully navigating the exponentially large configuration space \cite{Kwon2022}. At the atomistic modeling level, Chapman and Ma introduced a machine-learned spin-lattice potential trained via Gaussian process regression (GPR). Their surrogate model reproduced defect-driven magnetic dynamics in iron with near first-principles accuracy but at a fraction of the computational cost \cite{Chapman2022}. Liang et al. demonstrated the use of probabilistic neural networks embedded in artificial pyrochlore, establishing a hardware-efficient route to robust deep neural network implementations \cite{Liang2025}. Complementing these advances, Neogi et al. developed deep generative learning models based on variational autoencoders (VAEs) to interpret magnetic force microscopy images, enabling automated discovery of frustration patterns directly from experimental data \cite{Neogi2025}.Together, these studies illustrate that the development of tailored ML architectures—ranging from CNNs and GPR-based potentials to probabilistic neurons and VAEs—has transformed the study of defect sensitive material systems.

In this work, we focus on a critical and largely unsolved challenge in frustrated magnetism, which is to quantitatively characterize crystalline defects that sensitively modify the magnetic properties when there exist competing interactions~\cite{ramirez1994strongly, Balents2010, lacroix2011introduction}. Despite tremendous efforts and rapid advancements in the search for materials with novel quantum magnetic states, nearly all experimental observations remain obscured by the ambiguous and often dominant influence of disorder. When competing exchange interactions suppress the intrinsic energy scale, even weak defect potentials can become prominent, perturbing the delicate balance among intrinsically degenerate or nearly degenerate ground states. Such defect-driven effects have been widely implicated in several flagship frustrated magnets, such as Yb$_2$Ti$_2$O$_7$~\cite{arpino2017impact}, 1T-TaS$_2$~\cite{murayama2020effect}, $\alpha-$RuCl$_3$~\cite{zhang2023sample}, YbMgGaO$_4$~\cite{zhu2017disorder, kimchi2018valence}, and ZnCu$_3$(OH)$_6$Cl$_2$~\cite{kimchi2018scaling}. The model system we chose is Ho$_2$Ti$_2$O$_7$, which contains local magnetic moments carried by Ho$^{3+}$ that form a pyrochlore lattice ~\cite{gardner2010magnetic}. The geometry of this lattice, together with effective ferromagnetic interactions between nearest neighbors, enforces strong magnetic frustration, causing the moments to adopt a “two-in, two-out” configuration on each tetrahedron, analogous to the proton disorder in water ice. This so-called spin-ice state suppresses conventional long-range magnetic order and gives rise to emergent excitations and collective behavior that make Ho$_2$Ti$_2$O$_7$ a paradigmatic platform for studying frustrated magnetism~\cite{Harris1997, ramirez1999zero, Bramwell2001, Morris2009, Fennell2009}. Most intriguingly, crystalline defects influence the spin-ice state not through its ground-state configuration but through its dynamics~\cite{sala2014vacancy, wang2021monopolar}, highlighting defects as a potentially effective tuning route for magnetic excitations and relaxation processes without disrupting the underlying topologically constrained spin-ice manifold.

A common approach to characterizing crystalline defects, such as vacancies, site-mixing, interstitial defects, etc., is to analyze X-ray diffraction (XRD) measurements with Rietveld refinement techniques~\cite{Hooda2017, Kumar2017, Hatnean2017, Salari2012}. However, this approach faces several fundamental challenges. First, for physical properties that are highly sensitive to crystalline imperfections, such as spin dynamics of spin-ice, Rietveld analysis is insensitive to defect concentrations at the relevant levels, which can be as low as $\sim0.1-1\%$ or below~\cite{sala2014vacancy, wang2021monopolar}. Second, limitations at the data-generation stage, including imperfect correction of absorption and extinction effects and the difficulty of reliably integrating weak diffraction peaks, further restrict sensitivity to subtle disorder. Third, crystalline defects frequently manifest as \textit{local} structural distortions or nanoscale phase intergrowth (e.g. pyrochlore-fluorite motifs~\cite{OQuinn2020predicting}) that explicitly break translational symmetry and therefore cannot be accurately captured by refinements assuming a single periodic unit cell. As a consequence, conventional XRD refinement yields an averaged, projected description of disorder, obscuring the local defect configurations that are most relevant for emergent magnetic properties.

\onecolumngrid 

\begin{figure}[htbp]
    \centering
    \includegraphics[width=\textwidth]{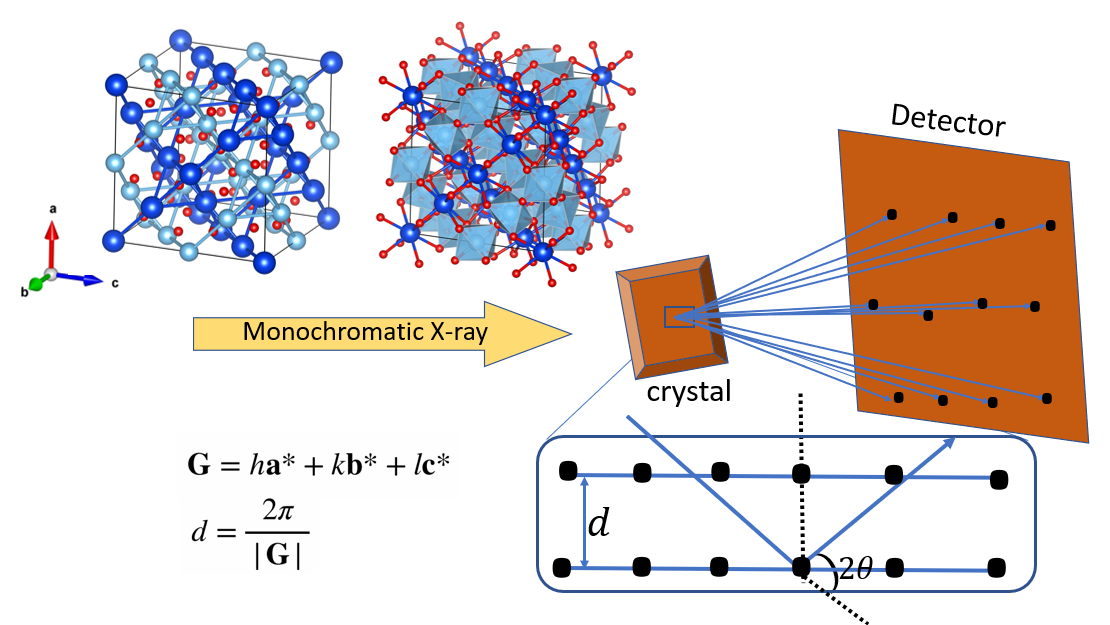}
    \caption{Schematic of the crystal structure, single-crystal X-ray diffraction geometry, and associated
    reciprocal-space construction used in this work. In the two crystal models shown as insets, Ho, Ti, and O atoms are represented by dark blue, light blue, and red balls, separately. The crystal model on the left highlights the two inter-penetrating tetrahedra networks formed by Ho-Ho and Ti-Ti connections, while the crystal model on the right highlights the oxygen environments and the TiO${}_6$ cage (light blue octahedra). The cubic lattice vectors are indicated by the $\bf a, b, c$ vectors forming the coordinate system shown as red, blue, and green arrows. A monochromatic X-ray beam
    illuminates the single crystal, generating Bragg-diffracted beams that intersect
    the area detector, where each spot corresponds to a reciprocal lattice vector
    $\mathbf{G} = h\mathbf{a}^* + k\mathbf{b}^* + l\mathbf{c}^*$. The magnitude of
    $\mathbf{G}$ determines the real-space lattice spacing
    $d = 2\pi / \lvert \mathbf{G} \rvert$. The inset at the bottom right illustrates the scattering plane,
    showing the relationship between the incident and diffracted beams, the scattering
    angle $2\theta$, and the projected interplanar spacing $d$.}
    \label{fig:Fig16}
\end{figure}

\twocolumngrid

In the current work, we focus on addressing the first challenge, namely, establishing a statistically robust refinement strategy that is fundamentally distinct from Rietveld analysis, while retaining standard community practices for data generation and correction. Motivated from \cite{Samarakoon2020}, we have expanded into a hybrid ML framework that integrates a pc-VAE with different BO methods to accelerate crystal structure refinement in pyrochlore system Ho$_2$Ti$_2$O$_7$. This pc-VAE projects the rough high-dimensional diffraction space to more continuous low-dimensional latent space first and then aims to reconstruct more realistic diffraction patterns, thereby providing a physically meaningful latent space. Then, pc-VAE assists traditional BO and high-dimensional Bayesian optimization (SAASBO) to efficiently explore the region of interest of low deviation between experimental XRD and GSAS-II simulated data. We have also explored over different parameter space to gain confidence in the refinement process. Finally, the solutions found via different approaches of autonomous exploration are reported and compared with traditional Rietveld refinement. This paper introduces a robust framework for applying advanced machine learning methods to enhance the precision of theoretical model refinement. By improving the accuracy of simulated diffraction data, it reduces reliance on expensive X-ray diffraction experiments. More accurate refinements, in turn, support the rapid generation of high-quality data, enabling faster discovery of new insights into the underlying physics of defect-sensitive, complex magnetic systems.


\section{Methods}

\subsection{\label{sec:level2}Crystal structure, diffraction data, and structure factor calculation of pyrochlore system Ho$_2$Ti$_2$O$_7$}
Ho$_2$Ti$_2$O$_7$ crystallizes in the pyrochlore structure with a cubic lattice (space group $Fd\bar{3}m$, No. 227), where Ho$^{3+}$ and Ti$^{4+}$ cations occupy the $16d$ and $16c$ Wyckoff sites, respectively, forming two interpenetrating networks of corner-sharing tetrahedra (\textbf{Fig.~\ref{fig:Fig16}}). The oxygen atoms reside in the $48f$ (O1) and $8b$ (O2) positions, coordinating the TiO$_6$ octahedra (\textbf{Fig.~\ref{fig:Fig16}}) and generating the characteristic pyrochlore A$_2$B$_2$O$_7$ framework. Structural parameters typically refined include atom positions that are not fixed by symmetry, the occupancy factor (Occ), which specifies the fraction of a given atomic site that is filled by the designated atom (Occ = 1.0 indicates a fully occupied site with no detectable vacancies or substitution), and the atomic displacement parameter (U$_{iso}$), which describes the mean-squared thermal motion of atoms around their average crystallographic positions. Together, these refined parameters characterize site disorders.

We collected x-ray diffraction data from a high-quality single crystal of Ho$_2$Ti$_2$O$_7$ grown using the traveling solvent floating zone (TSFZ) technique~\cite{ghasemi2018pyrochlore}. X-ray diffraction data were collected on a crystal of $0.1\times0.1\times0.1$ mm$^3$ using a SuperNova (Mo) micro-focus sealed-tube X-ray source ($\lambda = 0.71073$~Å) and a four-circle diffractometer equipped with an Atlas CCD detector. Measurements were performed at $T=110(2)$~K under a nitrogen atmosphere using $\omega$ scans to ensure full reciprocal-space coverage. Diffraction intensities were recorded over the range $3.5^\circ \leq\theta\leq36.1^\circ$. Absorption corrections were applied using analytical and spherical-harmonics–based scaling procedures implemented in CrysAlisPro. The collected dataset was then used for cell refinement and structural characterization to generalize the .hkl file that reports diffraction intensities $I_{hkl}$ at diffraction wavevector ${\bf G} = ha^*+kb^*+lc^*$ with $a^*, b^*, c^*$ the reciprocal lattice vectors (\textbf{Fig.~\ref{fig:Fig16}}).

X-ray diffraction intensity is directly connected to the structure factor $F_{hkl}$ by
\begin{equation}
    I_{hkl}=N \times |F_{hkl}|^2 \label{intensity-SF}
\end{equation}
where $N$ is a normalization factor determined by sample mass and instrument geometry, and $F_{hkl}$ is calculated from the microscopic crystalline model by
\begin{equation}
    F_{hkl} = \sum^N_{j=1} f_j \times e^{[2\pi i(hx_j + ky_j +lz_j]} \label{Eg. 11}
\end{equation}
Here, $f_j$ is the scattering factor of the j\textit{th} atom located at position $(x_j, y_j, z_j)$, which are atomic coordinates. \textbf{For simplicity in notation, we use $F$ to refer to  $|F_{hkl}|^2$ from here forward}. 

Single-crystal diffraction data differ fundamentally from powder diffraction data in both data structure and physical relevance. While powder diffraction reduces scattering information to a one-dimensional function of the diffraction angle $2\theta$ (\textbf{Fig.~\ref{fig:Fig16}}) through orientational averaging, single-crystal diffraction probes the structure factor defined on a three-dimensional reciprocal lattice, with intensities indexed by reciprocal-space coordinates $(hkl)$ ($\bf G$ vectors in \textbf{Fig.~\ref{fig:Fig16}}). Unlike high-dimensional image-based datasets such as diffuse neutron scattering spectra~\cite{Samarakoon2020}, where intensity is distributed continuously over reciprocal space, single-crystal diffraction data are intrinsically sparse: meaningful intensities occur only at discrete integer reciprocal lattice points satisfying the Bragg condition, while the vast majority of reciprocal-space manifest zero intensity. To reflect this sparsity and preserve physical meaning, we organize the three-dimensional single-crystal diffraction data as an ordered one-dimensional list indexed by $(hkl)$, creating a data structure that is computationally simple, analogous to powder diffraction, yet fundamentally distinct in that each data point corresponds to a specific location in three-dimensional reciprocal space rather than a powder-averaged scattering angle.

\textbf{Eqs.~\ref{intensity-SF}} and \textbf{~\ref{Eg. 11}} are used for the conventional Rietveld refinement in many GUI-based software packages and are employed in the current work to generate training data sets. Here, we chose the GSAS-II~\cite{toby2013}  package to provide benchmarking Rietveld refinement results and to simulate structure factors, $F$, for varying crystal parameters. The main advantage is that the scripting interface (GSASIIscriptable)~\cite{ODonnell2018Scriptable}, available in GSAS-II allows a seamless integration of established and newly-developed methods, but similar non-GUI access is not available in many other packages. 

\subsection{\label{sec:level2}Physically-Constrained Variational Autoencoder (pc-VAE):}

A variational autoencoder is a deep generative probabilistic model that belongs to the family of probabilistic graphical models and variational Bayesian methods. The two primary components of any VAE models are the encoder and decoder. Here, XRD data, a collection of thousands of \textit{apparently independent} values of $F$ corresponding to ${hkl}$ coordinates, represents a high-dimensional input; the encoder projects such an input into a low-dimensional latent variables, Z, following a distribution $p(Z|F)$. Then, with probabilistic sampling in the latent space (generally assumed as normally distributed), the decoder reconstructs the low dimensional latent variables into the respective estimated high-dimensional diffraction, $\overline{F}$,  following a distribution $p(\overline{F}|Z)$. Here, we designed the encoder and decoder with two fully connected neural networks, each of them having 4 hidden layers. This encoding-decoding process of the VAE models needs to be optimized such that the models can best learn the training data with minimum loss of information. The loss function of a VAE at each training epoch, $l_{vae}$, which we minimize, can be defined as the sum of reconstruction error ($\varphi$) and Kullback–Leibler (KL) divergence ($D_{KL}$), which can be mathematically written as \textbf{Eq. \ref{eq:vaeloss}}:
\begin{equation}
l_{vae}=\varphi + \beta \times D_{KL}(p(Z|F)||p(Z)) \label{eq:vaeloss}
\end{equation}
Here, the reconstruction error $\varphi$ can be chosen as the mean square error between the training input XRD and the reconstructed diffraction. The KL divergence $D_{KL}(p(Z|F)||p(Z))$ is the distance loss between the prior $p(Z)$ distribution (usually chosen as standard Gaussian) and the posterior $p(Z|F)$ distributions of the latent representation from data, while $\beta$ is the continuous scale factor at each training epoch. In this case, we have considered the default constant value of $\beta=1$. VAE models have been implemented to materials systems on various tasks like classification, feature or pattern recognition, prediction, etc. through unsupervised, semi-, or supervised learning.

From \textbf{Eq.~\ref{intensity-SF}}, it is apparent that x-ray scattering intensity differs from theoretically calculated $F$ by an experiment-related constant, determined by sample size and alignment, illumination volume, detector efficiency, etc. As the VAE can only be trained according to the model-predicted structure factors, $F$, a scale-invariance constraint has to be imposed during the VAE training. To isolate the defect-induced modifications to structure factors from global scaling of scattering intensities, diffraction data are normalized by the $(222)$ peak prior to VAE training. Here, the diffraction order $(222)$ was chosen because it represents one of the strongest and most reproducible reflections in our system — hence relatively stable against modest stoichiometric variations — making it a robust internal reference across both training and experimental datasets. This strategy mirrors practices in recent ML-based diffraction studies where “intensity rescaling” is employed \citep{PhysRevB.99.245120, Li2024AccurateMicroXRDMTL, choudhary2025}. As a result, the VAE is driven to capture physically meaningful variations linked to crystalline defects. Out of total $N$ sets of XRD data, thus, for the $j$th peak out of the total $J$ peaks in the $n$th data, the $F_{n,(j:hkl)}$ is first scaled with the respective $F_{n,(222)}$, followed by a multiplication with $\Psi = 1000$ to avoid numerical issues of rounding and digitization. Then, the data eventually passed into the VAE training process contain the scaled structure factor $F^s$ calculated as \textbf{Eq.~\ref{eq:F_scaling}}:
\begin{equation}
F^s_{n,j}=\frac{F_{n,j} \times \Psi}{F_{n,(222)}} \label{eq:F_scaling}
\end{equation}

Another critical aspect of the training of the unsupervised VAE model is to ensure the physical relevance of the latent space and can be validated from the reconstruction of the training data. In this case, the intensities of the XRD are always non-negative. In traditional VAE, this physical information is not known where the VAE solely focuses on minimizing the data loss. To mitigate this, we integrated the physical information as an absolute value constraint in the decoder where at every epoch in the training process, the reconstructed diffraction $\overline{F^s}$ is validated and transformed into the absolute value as $|\overline{F^s}|$. The reconstruction error $\varphi$ in \textbf{Eq.~\ref{eq:vaeloss}} with the stated physical constraint (pc) imposed thus became $\varphi_{pc}$, which is computed as \textbf{Eq.~\ref{eq:pc_ReconErr}}
\begin{equation}
\varphi_{pc}= \sum_n^N\sum_j^J (F^s_{n,j} - |\overline{F^s_{n,j}}|)^2 \label{eq:pc_ReconErr}
\end{equation}
As clearly seen from above, the physical constraint of positive semi-definiteness was essentially imposed in the way that the loss function value increases for an increased physical violation. Following \textbf{Eq.~\ref{eq:vaeloss}}, the loss function of the scaling invariant pc-VAE at each training epoch, $l_{pc-vae}$, can be modified as \textbf{Eq.~\ref{eq:pc_vaeloss}}:
\begin{equation}
l_{pc-vae}=\varphi_{pc} + \beta \times D_{KL}(p(Z|F^s)||p(Z)) \label{eq:pc_vaeloss}
\end{equation}

\begin{figure}
    \centering
    \includegraphics[width=1\linewidth]{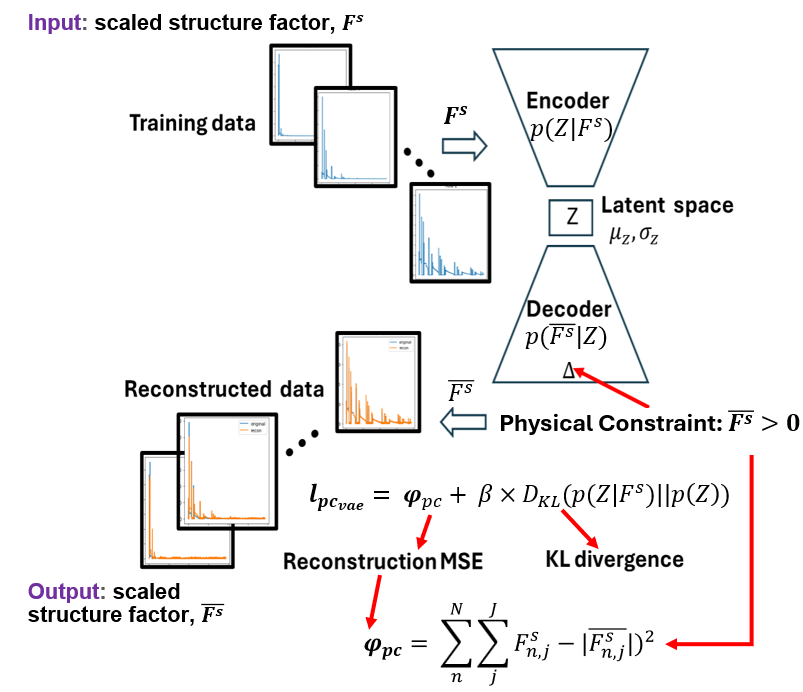}
    \caption{Proposed architecture of the pc-VAE with physical constraints and scale invariance. Here, the physical constrained is defined as the intensity of the reconstructed diffraction are non-negative.}
    \label{fig:1}
\end{figure}

The necessities of imposing the scaling-invariance and positive semi-definiteness will be demonstrated in details in the Results section. In \textbf{Fig.~\ref{fig:1}}, we show the architecture of our VAE with the physical constraints imposed, \textit{i.e.}, pc-VAE. Previously, VAE has been utilized for dimension reduction and key feature extractions from high-D XRD data \cite{Banko2021, Yamashita2022} where improvement in maximizing extraction of physically-relevant information from material systems has been achieved via integrating physical bias \cite{Biswas2024pVAE}.

\subsection{\label{sec:level2}Bayesian Optimization (BO) and Sparse Axis-Aligned Subspace BO (SAASBO):}

\begin{figure}
    \centering
    \includegraphics[width=1\linewidth]{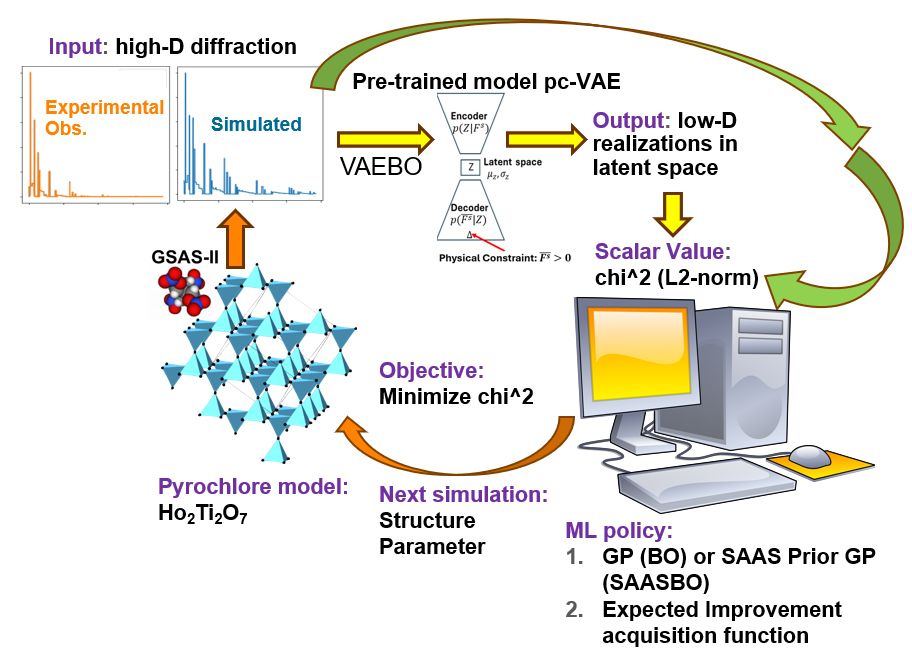}
    \caption{Proposed architecture of the pc-VAE assisted BO and SAASBO exploration for crystalline structure refinement of pyrochlore model Ho$_2$Ti$_2$O$_7$. The yellow arrows are the additional steps for the pc-VAE BO while the green arrows are the steps for traditional BO. The orange arrows are the common steps for the BO and the pc-VAE-BO.}
    \label{fig:2}
\end{figure}

\textbf{Fig. \ref{fig:2}} shows the workflows of the traditional and proposed pc-VAE assisted BO-driven exploration. Once we developed the pretrained pc-VAE model, the next task is to couple with the BO framework to undergo autonomous refinement of the crystalline structure parameters of the pyrochlore model. Here, starting with a few initial samples of structure parameters (either generated randomly or combining with domain knowledge), we compute the high-D simulated diffraction data and the experimental observation (from GSAS-II). This high-D data structure is then projected into the low-D latent space of pretrained pc-VAE ($\Delta$). Then, the L2-norm between the latent representation of the simulated ($F^s_{cal,l}$) and experimental ($F^s_{obs,l}$) diffraction, notated as $\chi_l^2$, is calculated over 10000 of Monte Carlo cycles as \textbf{Eq.~\ref{eq:Chisqlatent}}:
\begin{align}
\chi^2_l = \sum^{10000}_{k=1}||F^s_{obs,l}(\Delta_k)-F^s_{cal,l}(\Delta_k)||_2 \label{eq:Chisqlatent}
\end{align}
To improve robustness in the objective function, we computed the mean of the L2-norm via Monte Carlo simulation. At each loop of the BO, given the current training samples of evaluated structure parameters with the respective $\chi_l^2$, a prediction model such as Gaussian Process (GP) \cite{Rasmussen:2005gp} is fitted to estimate the $\chi_l^2$ of all the unevaluated structure parameter samples. Then, the acquisition function of BO such as Expected Improvement \cite{Brochu2010} is maximized (via maximizing the negative $-\chi_l^2$) to suggest the next structure parameter samples for GSAS-II simulation. This loop is continued, where GP is updated iteratively with data augmentation, till the optimal solution is found or the exploration cost is exhausted. Previously GP and BO has been extensively used for accelerated and efficient exploration over various time-expensive continuum \cite{Morozovska2021, Morozovska2022}, classical \cite{Ziatdinov2022, Valleti2022, Biswas2024iMFBO} and quantum simulation models \cite{Samarakoon2020, Thamm:2025bk} of material systems, to identify the optimal conditions.

Here, we employ a surrogate non-parametric Gaussian process (GP) model \cite{Rasmussen:2005gp} defined as below \textbf{Eqs.~\ref{eq.5}} and \textbf{~\ref{eq.6}}. This GP model was fitted using the \texttt{GPyTorch} \cite{Gardner:2018gp} and \texttt{BoTorch} \cite{Balandat:2020bt} software libraries.
\begin{align}
y(\mathbf{x}) &=\mathbf{x}^T \cdot \beta + z(\mathbf{x}), \label{eq.5}\\
z(\mathbf{x}) &\sim \mathrm{GP}\qty(\mathbb{E}[z(\mathbf{x})], \mathrm{cov}(x,x^{\prime})) \label{eq.6}
\end{align}
Here, $y(\mathbf{x})$ is the estimation of $y$ given input $\mathbf{x}$, $\mathbf{x}^T \cdot \beta$ is a user chosen polynomial regression model. In this paper, we have considered Constant Mean model with the tunable hyperparameter of the constant value, $\mu(x)=C$. $z(\mathbf{x})$ is the Gaussian process regression with zero mean, \textit{i.e.}, $\mathbb{E}[z(\mathbf{x})]=0$. The covariance function $\mathrm{cov}(x,x^{\prime})$ between inputs $x$ and $x^\prime$ is computed using a Piecewise-Polynomial kernel function as \textbf{Eqs.~\ref{eq.7}, ~\ref{eq.8}} and \textbf{~\ref{eq.9}}:
\begin{multline}
        K_{ppD,2}(x, x^\prime) = (1 - r)^{(j+2)} \\
        + \left(1 + (j+2)r + \frac{j^2 + 4j + 3}{3}r^2\right)
        \label{eq.7}
\end{multline}
where
\begin{align}
    r &= \frac{\lVert x - x^\prime \rVert}{\theta}, \label{eq.8}\\
    j &= \left\lfloor \frac{D}{2} \right\rfloor + q + 1 \label{eq.9}
\end{align}
Here, $D$ is the dimension of the input $\textbf{x}$ and $q$ is the smoothing parameter which is set as $q=2$, as per the default suggestion in \texttt{GPyTorch}. The hyper-parameter $\theta$, representing the length scale, is optimized via a gradient based method such as the Adam optimizer \cite{Kingma2014} with learning rate $=0.1$ and weight decay $=0.01$. The Expected Improvement (EI) acquisition function takes the following mathematical form, given $\sigma(x)^2 > 0$, as \textbf{Eq.~\ref{eq.EIacqfunc}}:
\begin{multline}
     \mathrm{EI}(x) = \qty(\mu(x)-y_{\rm best}-\epsilon) \times \Phi\qty(\frac{\mu(x)-y_{\rm best}-\epsilon}{\sigma(x)})+ \\
     \sigma(x) \times \phi\qty(\frac{\mu(x)-y_{\rm best}-\epsilon}{\sigma(x)})
     \label{eq.EIacqfunc}
\end{multline}
Here, $\mu_y(x)$ and $\sigma_y(x)$ are the GP predictive mean and standard deviation of the $\chi_l^2$, $y_{\rm best}$ is the best value of the $\chi_l^2$ in the current training set, $\Phi(\cdot)$ is the cumulative normal distribution function, and $\phi(\cdot)$ is the normal probability distribution function. A slack parameter $\epsilon = 0.01$ was added for numerical stability and to balance between exploration and exploitation \cite{Jones2001}. For any $\sigma(x)^2 = 0$, $EI(x) = 0$.

However, the performance of the traditional BO is limited to low-dimensional exploration and increases challenges for true convergence as the dimension of the parameter space increases.  Methods have been attempted to tackle BO in high dimensional problems through a different strategy of projection with random embedding and quantile Gaussian Process to a reduced space \cite{Wang2016, Moriconi2020} or projecting to a latent space \cite{Biswas2024zBO}, or using special kernels \cite{Oh2018}. As one of the critical objectives is to obtain the optimal solutions with high precision, accuracy and physical relevance, we have also considered the above defined autonomous refinement with Sparse Axis-Aligned Subspace BO (SAASBO) \cite{Eriksson2021} in \texttt{BoTorch}. SAASBO is specifically designed for high-dimensional Bayesian optimization where it leverages a sparse prior on the inverse lengthscales of a Gaussian Process (GP) kernel—known as the SAAS prior—to estimate the irrelevant structure parameters (i.e., assigned large length-scales) while allowing other structure parameters to significantly influence the objective space. This approach enables automatic relevance determination without requiring prior knowledge of which variables matter. Posterior inference is performed using Hamiltonian Monte Carlo (HMC) to sample from the joint posterior over GP hyperparameters and latent functions, ensuring robust uncertainty quantification. Then, we employ the similar Expected Improvement acquisition function to propose new input points. By iteratively refining the model and focusing the search within a learned low-dimensional subspace, SAASBO achieves sample-efficient optimization in problems with many irrelevant variables. As in \textbf{Fig. \ref{fig:2}}, to convert from BO to SAASBO, we have only replaced the prediction model in the ML Policy with the SAAS Prior-Gaussian Process (SAASGP). To summarize, we explored the structure parameter space of the pyrochlore Model via 1) BO with first fixing 4 most relevant parameters from domain informed knowledge, 2) BO with further dimension reduction with iteratively learned significant parameters for refined exploration and 3) SAASBO with considering all 8 parameters to further validate the optimal solutions.

\section{RESULTS}
In this section, we have provided the performance of the pc-VAE trained with scaled XRD data, followed by reporting of the converged solutions of different autonomous exploration via pc-VAE assisted BO and SAASBO.

\subsection{\label{sec:level2}Physically-Constrained Variational Autoencoder (pc-VAE):} To train the pc-VAE, we first generated 25000 GSAS-II simulated diffraction patterns of the pyrochlore model Ho$_2$Ti$_2$O$_7$ (following \textbf{Eq. \ref{Eg. 11}}) over the structure parameter space. The structure parameters not fixed by the space group are: fractional $x$-coordinate of the first oxygen site with Wyckoff position of $48f$ (\textbf{xO1}), occupancy of the second oxygen site with Wyckoff position of $8b$ (\textbf{OccO2}), atomic displacements for both oxygen atoms (\textbf{UO1, UO2}), site occupancy and displacement parameters for Ti (\textbf{OccTi, UTi}) and Ho (\textbf{OccHo, UHo}). The displacement terms have units of \AA$^2$ while the other parameters are unit-less. Here we fix the occupancy of O1 (\textbf{OccO1}) at 1 to remove the redundant degree of freedom in the overall scaling factor. Over this 8-D parameter space, 25000 training samples were generated using the Latin Hypercube sampling method to improve the coverage of the large space than using randomly generated samples. Each of the 25000 diffraction patterns comprises 2569 unique diffraction orders indexed as $(hkl)$. The 25000 simulated patterns were organized into a data frame with the diffraction patterns corresponding to different structural parameters as rows and Miller indices as columns, and this data frame was fed into the VAE training pipeline.

\begin{figure}[htbp]
    \centering
    \includegraphics[width=1\linewidth]{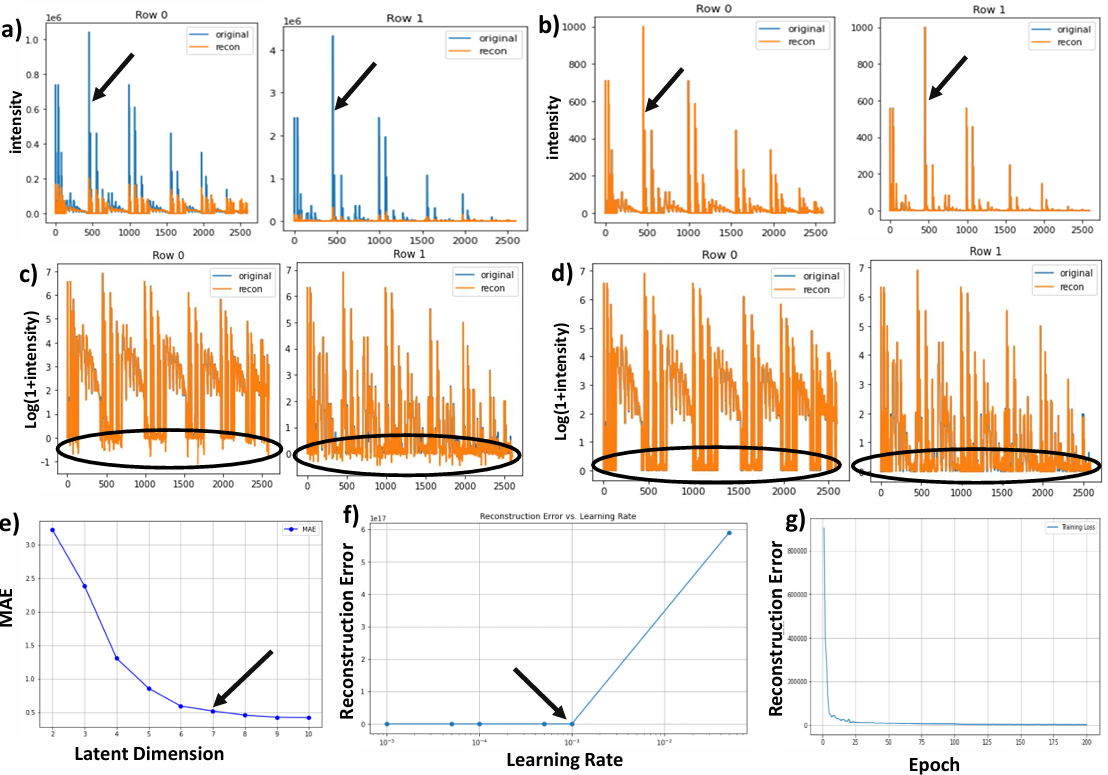}
    \caption{Performance of the implementation of scaling invariance, physical constraints and hyperparameter optimization of pc-VAE. Panels (a) and (b): The performance comparison of VAE without and with the scaling invariance respectively. No physical constraints had been imposed in this case. Panels (c) and (d): The performance comparison for VAE trained without and with the physical constraint for non-negativity of diffraction intensities respectively. No scaling invariance have been imposed in this case. The black circled region highlights the physical constraint violation of the reconstruction in (c) and no such violation in (d). For panels (a)-(d), two examples of reconstruction of the diffraction pattern have been shown out of 25000 training data. Panels (e), (f) and (g): Optimized hyperparameters such as latent dimension, learning rate and training epochs respectively, for the training process of pc-VAE with imposing scaling invariance and physical constraints.}
    \label{fig:3}
\end{figure}

\textbf{Fig.~\ref{fig:3}} shows the performance comparison between proposed pc-VAE and traditional VAE without imposing scaling invariance and physical constraints, and the training process of pc-VAE with hyperparameter optimization. Firstly, we have trained the network on both scaled and unscaled datasets. The scaling was done following \textbf{Eq.~\ref{eq:F_scaling}}. It is clearly demonstrated in \textbf{Fig.~\ref{fig:3}a-b} that the scaling significantly improved the performance of the model in predicting the peak intensities. The reconstruction root mean square error (RMSE ) achieved by training on scaled data is $26.06$, while the model trained on unscaled data reached a RMSE  of $185.69$ at its best performance. The diffraction patterns used for the training contain a collection of diffraction intensities that are very weak, corresponding to symmetry-allowed diffraction orders, but with very small structure factors. Traditional VAE often reconstructs these zero values as negative values (inside the black circle in \textbf{Fig. ~\ref{fig:3}c}) as compared to our pc-VAE (\textbf{Fig. ~\ref{fig:3}d}), highlighting the necessity to impose the physical constraint to keep the model physically meaningful. Several options were attempted to impose this physical constraint, with the best being to introduce an absolute value function into the network. This absolute value function acts on reconstructed patterns at each epoch before the calculation of reconstruction loss. With this, the absolute value function became part of the network (Figure~\ref{fig:1}). In addition to making the model physically meaningful, the physical constraint also improved the performance of the model from a RMSE of 30.77 to 26.06. To train the pc-VAE, we started by first optimizing the hyperparameters, specifically, latent dimension, learning rate, number of epochs, and the network structure. To identify the optimal latent dimension, we trained the pc-VAE for latent dimensions ranging from 2 to 10, and the performance of the model improved with increases in latent dimension from latent dimension of 2 to 7. After the latent dimension of 7, further improvement in model performance was no longer significant. Hence, the latent dimension of 7 was selected as the optimal latent dimension as illustrated in \textbf{Fig.~\ref{fig:3}e}. As seen in Fig.~\ref{fig:3}f, the pc-VAE was also trained for learning rates ranging from $10^{-5}$ to $10^{-1}$ and the optimal learning rate, considering both reconstruction loss and model complexity, is $10^{-3}$. Using the optimal hyperparameters above, the model was trained for 200 epochs, and it was observed (\textbf{Fig.~\ref{fig:3}g}) that the model converged at epoch of 100. In addition to the above, different network structures were also tried, and the network structure with the best performance was used in training the final model. 

\newpage
\onecolumngrid 

\twocolumngrid

\subsection{\label{sec:level2}Structure Parameter Refinement of Ho$_2$Ti$_2$O$_7$:}

In this section, we provide detailed analysis of the convergence of the refinement of the structure parameters via traditional BO, SAASBO, pc-VAE BO, and pc-VAE SAASBO. We started with the domain-informed confined parameter settings as \textbf{$0.35 \leq$ xO1 $\leq 0.45$, $0.9 \leq $OccTi, OccHo, Occ02$ \leq 1.1$ and $0 \leq $UTi, UHo, UO1, UO2$ \leq 0.1$ \AA$^2$}; with \textbf{OccO1} = $1$ fixed as explained earlier. \textbf{Table \ref{tab:refined_params_Domain}} provides the summary of the optimal solutions from different BO approaches. We note that none of the refined crystal structure parameters represent ground-truth knowledge and we mainly focus on the statistical robustness within current model. Various factors that affect the accuracy of the refinement, such as extinction correction and local defects, were not considered in the current work. Therefore, we will focus on the statistical manifestations in the parameter space and the indicators $\chi^2$ in the presentation of results.

Firstly, we conducted and reported the results from 4-D space exploration via traditional BO and pc-VAE BO. Based on the domain expert knowledge about the physical importance of the parameters, the 4 preferred structure parameters chosen for refinement are site occupancy and displacement for Ti and Ho atoms (i.e. \textbf{OccTi, UTi, OccHo, UHo}). Based on the initial structural refinement using the conventional Rietveld method in GSAS-II, the other 4 parameters are fixed at the values obtained from Rietveld refinement such as xO1=0.42176, OccO2=0.9204, UO1=0.0083, UO2=0.00325. The initial guess for the domain-expert chosen control structure parameters are OccTi=0.9, UTi=0.0, OccHo=0.9173, UHo=0.0033, as similarly obtained from the Rietveld refinement. The $\chi^2$ values of the Rietveld refined parameters, computed in the real space (high-D) and the latent space of the pc-VAE (low-D) are $\chi^2 = 30.52$ and $\chi^2_l = 3.2259$ respectively. Our goal is to find better solutions that can further reduce the value of $\chi^2$ and/or $\chi^2_l$.

\newpage
\onecolumngrid
\newcommand{\B}[1]{\textbf{#1}} 
\newcommand{\R}[1]{#1}          

\begin{table}[htbp]
\centering
\caption{\textit{Summary of results of structure refinement from different exploration strategies with domain-informed parameter space.}}
\label{tab:refined_params_Domain}
\begin{tabular}{lrrrrrrrrrr}
\toprule
Analysis Options & Xo1 & OccO2 & UO1 & UO2 & OCCTi & UTi & OccHo & UHO & $\chi^2$ & $\chi^2_{l}$ \\
\midrule

Rietveld Refinement      & \textit{\color{brown}0.4216} & \textit{\color{brown}0.9204} & \textit{\color{brown}0.0083} & \textit{\color{brown}0.00325} & \textit{\color{brown}0.9}& \textit{\color{brown}0.0}    & \textit{\color{brown}0.9173}& \textit{\color{brown}0.0033} & \textit{\color{brown}30.52}   & \textit{\color{brown}3.2259} \\

4-D BO            & 0.421597 & 0.920386 & 0.008262 & 0.00325 & \B{1.0}     & \B{0.0}    & \B{0.92}     & \B{0.01}     & \color{red}\B{20.4004}   &  4.7750\\
2-D BO            & 0.421597 & 0.920386 & 0.008262 & 0.00325 & \B{0.9545}     & 0    & \B{0.9263}  & 0.01   & \B{20.4029} &  4.6674\\
4-D VAE-BO          & 0.421597 & 0.920386 & 0.008262 & 0.00325 & \B{0.9}     & \B{0.0}    & \B{1.06}     & \B{0.0}     & 58.3803 & \B{2.5189} \\
2-D VAE-BO          & 0.421597 & 0.920386 & 0.008262 & 0.00325 & \B{0.9}  & 0    & \B{1.0515}  & 0     & 57.505 & \color{red}\B{2.5170} \\
8-D SAASBO    & \B{0.421597} & \B{0.920386}   & \B{0.008262}  & \B{0.00325}  & \B{1.0}  & \B{0.0} & \B{0.92}  & \B{0.01} & \B{20.4004}  & 4.7750 \\
8-D VAE-SAASBO      & \B{0.421597} & \B{0.920386}   & \B{0.008262}  & \B{0.00325}  & \B{0.9}  & \B{0.0} & \B{1.0105}  & \B{0.0} & 57.505  & \B{2.5170} \\

\bottomrule
\end{tabular}
\medskip

\textit{Note}. Under each scenario, the bold values under the parameter columns are optimized while the non-bold are fixed in the setting, whereas the bold values under either the $\chi^2$ or $\chi_l^2$ columns are the one minimized in that exploration.
\end{table}



\begin{figure}[htbp]
    \centering
    \includegraphics[width=0.85\linewidth]{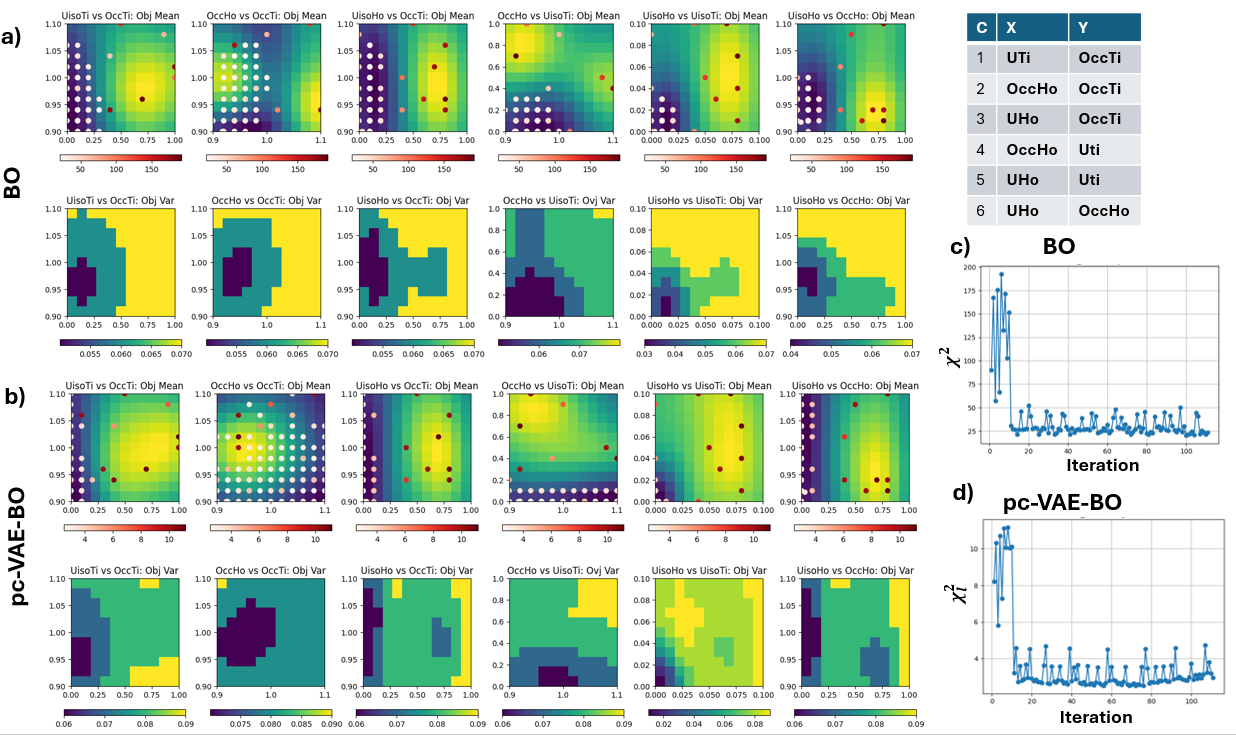}
    \caption{Convergence of BO and pc-VAE-BO over the 4-D structure parameter space. In each of the panels (a) and (b), top and bottom rows represent the maps of GP mean and GP uncertainty respectively, while panels (a) and (b) represent BO and pc-VAE-BO, respectively. The table on the top right indicates the structure parameters placed on the X and Y axes for each of the 6 columns in panels (a) and (b). Panels (c) and (d) present the convergence plots of $\chi^2$ and $\chi_l^2$ for BO and pc-VAE-BO respectively. }
    \label{fig:4}
\end{figure}

\twocolumngrid 

\textbf{Fig. \ref{fig:4}} shows the convergence of BO and pc-VAE-BO over the physically feasible 4D parameter space. Here, we have considered 10 randomly generated starting samples and 1 sample obtained from Rietveld refinement. Then, each parameter space is discretized into 11 samples, making total of $11^4 = 14641$ samples to choose for autonomous sampling. The total number of iterations is set to 100. On average, in each 4D exploration, the time taken for the completion of 100 iterations is about 3.5 hours. The machine specification conducted in this and the rest of all the reported explorations are: CPU with 16 GB RAM and Core i7 processor. Comparing the GP mean maps [top \textbf{figs. \ref{fig:4}(a) and (b)}], we can clearly see that the mean maps suggest the structure parameters \textbf{OccTi,  OccHo} are highly significant in the exploration, whereas the structure parameters \textbf{UTi, UHo} have minimal significance. Comparing figures (c) and (d), the pc-VAE-BO identifies the optimal region much quicker (after approximately iteration 70) than BO (after approximately iteration 100). This signifies that pc-VAE helped to improve the exploration to accelerate towards the optimal region, which suggests the significance of implementation of pc-VAE to improve the function space from noise and artifact solutions. Finally, the optimal solutions found from BO and pc-VAE-BO are OccTi=1.0, UTi=0.0, OccHo=0.92, UHo=0.01; $\chi^2$ = 20.4004 (objective function), $\chi_l^2$ = 4.7750 and OccTi=0.9, UTi=0.0, OccHo=1.06, UHo=0.0; $\chi^2$ = 58.3803, $\chi_l^2$ = 2.5189 (objective function). We can see the BO and pc-VAE BO provide the optimal solution with better $\chi^2$ and $\chi_l^2$, respectively, than that for Rietveld refinement. 
\newpage
\onecolumngrid 

\begin{figure}[htbp]
    \centering
    \includegraphics[width=0.8\linewidth]{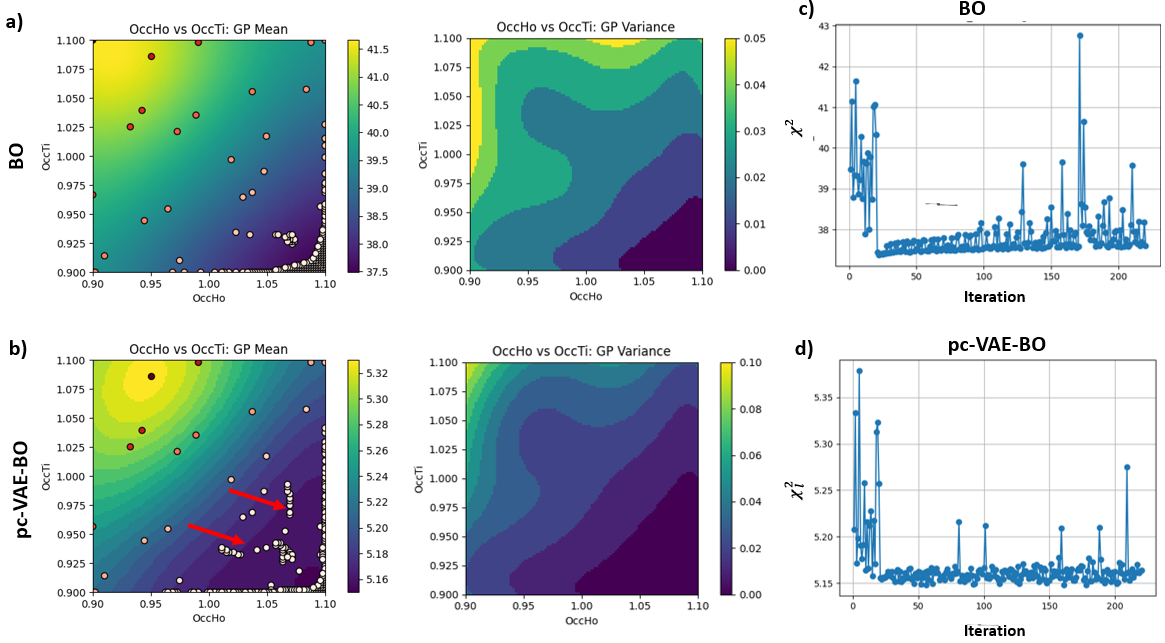}

    \caption{Convergence of BO and pc-VAE-BO over the 2-D structure parameter space, as selected based on the interpretation from Fig~\ref{fig:4}. In each of the panels (a) and (b), the first column represents the map of GP mean while the second column represents the GP uncertainty. Panels (c) and (d) present the convergence plots of $\chi^2$ and $\chi^2_l$ for BO and pc-VAE-BO respectively. }
    \label{fig:5}
\end{figure}

\twocolumngrid 

To refine the structure parameter space even further, based on the results in \textbf{figs. \ref{fig:4}a and b} and Table \ref{tab:refined_params_Domain}, we ignored \textbf{UTi, UHo}) in the exploration (fixed at UTi=0.0, UHo=0.01 for BO and fixed UTi=0.0, UHo=0.0 for pc-VAE-BO) and only considered \textbf{OccTi,  OccHo} in this analysis. As before, the other fixed structure parameters are at xO1=0.42176, OccO2=0.9204, UO1=0.0083, UO2=0.00325. Here, each parameter space is discretized into 100 samples, making total of $100^2$ = 10000 samples to choose for autonomous sampling. Here, we have considered 20 randomly generated starting samples. The total number of iterations is set to 200. To avoid getting trapped in a local minima, we avoided the samples from the initial guess in all the 2D analysis. \textbf{Fig. \ref{fig:5}} shows the convergence of BO and pc-VAE-BO over the stated 2-D parameter space. On average, in each 2D exploration, the time taken for the completion of 100 iterations is about 5 hours. Comparing the GP mean maps [left \textbf{figs. \ref{fig:5}(a) and (b)},] we can clearly see that the region of interest (dark region) of the pc-VAE-BO and the BO have started to deviate. However, we can see while BO predicts a wider region of interest (dark area) within the physical domain boundaries of the controlled structure parameters, pc-VAE-BO provides much confined region of interest, reducing noise and artifact solutions. Also interestingly, we can see the GP mean from BO signifies that \textbf{OccHo} is a non-significant structure parameter, deviating the analysis from the respective 4D analysis as in \textbf{Fig. \ref{fig:4}a}. On the other hand, we see the GP mean from pc-VAE-BO still suggests \textbf{OccHo} is a significant parameter, aligning with the results from the respective 4D analysis as in \textbf{Fig. \ref{fig:4}b}. This signifies the robustness of the exploration of pc-VAE-BO rather than BO due to random noise reduction in the latent function space. Figures (c) and (d) are the convergence plots for BO and pc-VAE-BO respectively, where now for less complex (2-D) parameter space, pc-VAE-BO converges even quicker than BO (highlighted by red arrows). Finally, the optimal solutions found from BO and pc-VAE-BO are OccTi=0.9545, OccHo=0.9263; $\chi^2$ = 20.4029 (objective function),$\chi^2_l$ = 4.6674 and OccTi=0.9, OccHo=1.0515; $\chi^2$ = 57.505, $\chi^2_l$ = 2.5170 (objective function). Similarly, we can see the BO and pc-VAE BO provides the optimal solution with better $\chi^2$ and $\chi^2_l$ respectively than that for Rietveld refinement.

\newpage

\onecolumngrid

\begin{figure}[htbp]
    \centering
    \includegraphics[width=1\linewidth]{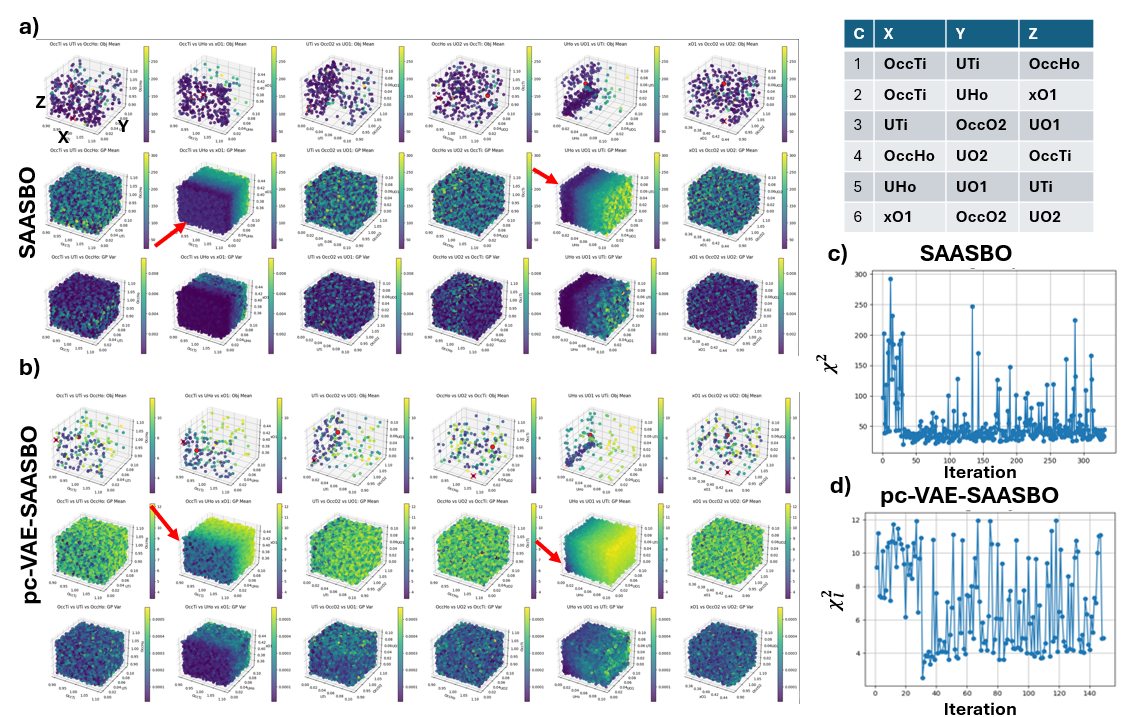}

    \caption{Convergence of SAASBO and pc-VAE-SAASBO over the broader 8-D structure parameter space. In each of the panels (a) and (b), top, middle and bottom rows represent the evaluated samples, maps of GP mean and GP uncertainty, respectively. The table on the top right indicates the structure parameters placed on the X and Y axes for each of the 6 columns in panels (a) and (b). Panels (c) and (d) present the convergence plots of $\chi^2$ and $\chi^2_l$ for SAASBO and pc-VAE-SAASBO, respectively. }

    \label{fig:6}
\end{figure}



\begin{figure}[htbp]
    \centering
    \includegraphics[width=0.8\linewidth]{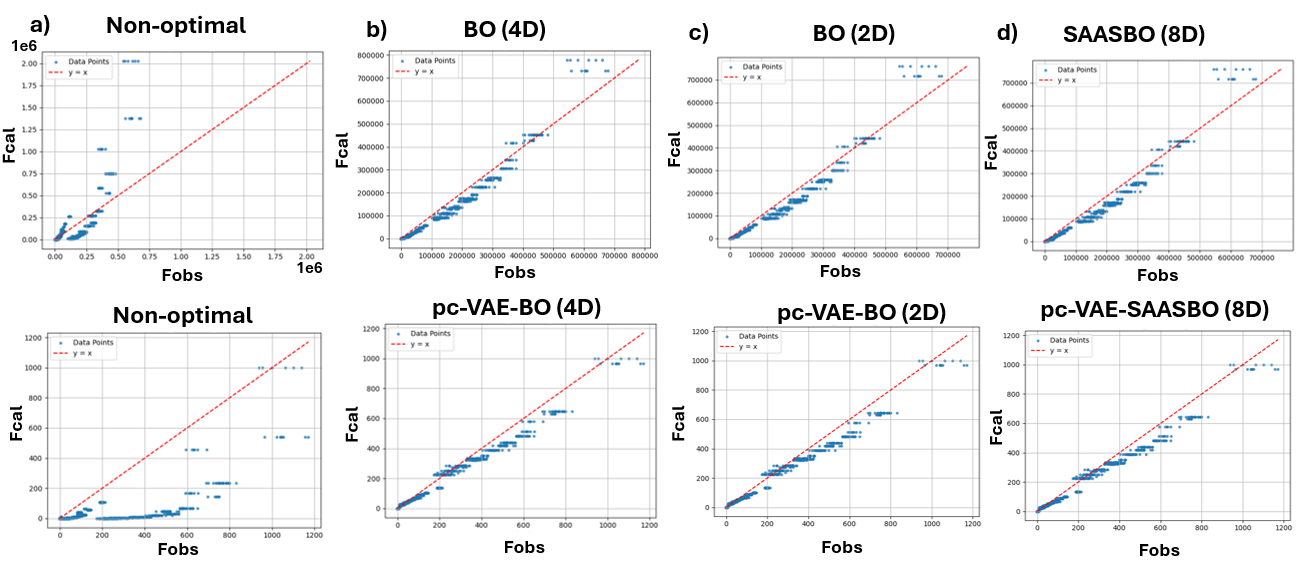}
    \caption{Comparison plots between experimental structure factor, $F_{obs}$, and simulated structure factor, $F_{cal}$, for the given structure parameter values over the crystal dataset. (a) is the plot for a random non-optimal structure parameter. The top figures of (b), (c) and (d) are the plots for optimal structure parameter values as provided in \textbf{Table \ref{tab:refined_params_Domain}} for 4D-BO, 2D-BO and 8D-SAASBO respectively. The bottom figures of (b), (c) and (d) are the plots for optimal structure parameter values as provided in \textbf{Table \ref{tab:refined_params_Domain}} for 4D-pcVAE-BO, 2D-pcVAE-BO and 8D-pcVAE-SAASBO respectively. For each plot, the red diagonal line $y=x$ referenced the degree of association between the experimental and simulated structure factors as the respective scatter plots aligns with it.}
    \label{fig:7}
\end{figure}
\twocolumngrid
Extending BO to SAASBO to allow exploring over 8-D structure parameter space, we considered exploration with only random starting samples, with Rietveld refined starting samples and with the starting samples obtained from the best solutions obtained between 2D and 4D explorations. Different explorations with different strategies of starting samples were done to validate extensively if we can obtain any better solutions that of the previous analysis. \textbf{Fig. \ref{fig:6}} shows the convergence of SAASBO and pc-VAE-SAASBO over the physically feasible parameter space. The details of the explored range of each structural parameter are provided in the figure captions. It is to be noted that, we reported the best optimal solution obtained from different initial sampling strategies as in Table \ref{tab:refined_params_Domain}. Within the stated ranges, we have considered 30 randomly generated starting samples, with further randomly generated 10000 samples to choose for autonomous sampling and the total number of iteration is set to 300. Comparison of the GP mean map [middle \textbf{figs. \ref{fig:6}(a) and (b)},] we can clearly see that region of interest (dark region) of the pc-VAE-SAASBO is more confined than the SAASBO, providing more robust region of interest. We know from earlier results that the optimal values of UTi and UHo are very close to zero. However, we can see from GP plot from SAASBO (\textbf{Fig. \ref{fig:6}a} 5th middle plot indicted by red arrow), the optimal region (dark region) is shown parallel to the axis of parameter UTi. On the other hand, the GP plot from pc-VAE-SAASBO (\textbf{Fig. \ref{fig:6}b} 5th middle plot indicted by red arrow) aligns with the previous interpretation where the optimal region (dark region) is predicted near UTi =0. From \textbf{figs. \ref{fig:6}(c) and (d)}, we can see that we did not find any better optimal solutions than the best solutions (highlighted in red in Table \ref{tab:refined_params_Domain}) used as starting samples. However, pc-VAE-SAASBO stopped as the maximum acquisition value approaches to zero, after approximately 120 iterations ($\approx$ 4 hours). Whereas, SAASBO still continue to explore redundantly as the acquisition function falsely guides to explore over noisy solutions, till the model stopped at the limit of 300 iterations ($\approx$ 8 hours). Finally, the optimal solutions found from SAASBO and pc-VAE-SAASBO are similar to the solutions obtained from 4D BO and 2D pc-VAE-BO analysis respectively. 

To understand the validation of the results as reported in \textbf{Table \ref{tab:refined_params_Domain}}, we have compared the alignment of the structure factors from the experiments $F_{obs}$ and the simulation model $F_{cal}$, as shown in \textbf{Fig. \ref{fig:7}}. For pc-VAE integrated explorations, we have plotted the scaled structure factor $F^s$ calculated per Eq.~\ref{eq:F_scaling}. This is intended to build a fair comparison of convergence as $F^s$ is what serves as the input to compute $\chi^2_l$. Comparing a random structure parameter (showing both unscaled and post-process scaled structure factors) in \textbf{fig. \ref{fig:7}a} with optimal structure parameters from different BO strategies in \textbf{figs. \ref{fig:7}b-d}, we can see the refinement of the simulation model to match with the experimental observations.

\section{CONCLUSION AND FUTURE TASKS}
To summarize, we have presented different exploration approaches via Bayesian optimization in order to improve precision of the refinement of the structure parameters of the pyrochlore model, Ho$_2$Ti$_2$O$_7$, to minimize the deviation of the structure factors between the XRD experimental observations and the calculated simulations. We collected x-ray diffraction data from a single-crystal specimen of Ho$_2$Ti$_2$O$_7$ grown from TSFZ growth. Initially, from 25000 simulated XRD datasets generated from Latin Hypercube sampling, a physically constrained Variational Autoencoder (pc-VAE) is developed and trained. We show the improvement of pc-VAE over traditional VAE in physical relevance and accuracy of data reconstruction. Then, this pc-VAE is integrated in the objective function space in autonomous exploration models such as BO and SAASBO to improve the model prediction and overall exploration and exploitation. Finally, various parameter spaces (2D, 4D and 8D) have been considered, via domain-informed knowledge regarding the importance or sensitivity to the pyrochlore model tuning. In general, we see the optimal solutions mostly lie near the edges of the physically-defined parameter spaces. In order to expand the parameter space, our future task would be to design the proposed architecture from purely data-driven acquisition function to developing domain-preferred cost-driven acquisition function where some local region of interest will be preferred more than the others, based on the physical relevance of the optimal solutions. Another future aim is to integrate the experimental Neutron diffraction data with XRD to explore in a multi-functional structure factor space of a pyrochlore model and discover the Pareto fronts in the structure parameter refinement via Multi-objective Bayesian optimization.      

\begin{acknowledgments}

This work (J.A) was supported by the University of Tennessee startup funding of A.B. The authors (J.A. and A.B.) acknowledge the use of facilities and instrumentation at the UT Knoxville Institute for Advanced Materials and Manufacturing (IAMM) and the Shull Wollan Center (SWC) supported in part by the National Science Foundation Materials Research Science and Engineering Center program through the UT Knoxville Center for Advanced Materials and Manufacturing (DMR-2309083). This research (Y.W and A.M) was partially supported by the National Science Foundation Materials Research Science and Engineering Center program through the UT Knoxville Center for Advanced Materials and Manufacturing (DMR-2309083). This work (H.D and Y.T) was performed during the Student Mentoring and Research Training (SMaRT) program, jointly supported by the U.S. Department of Energy's Office of Energy Efficiency and Renewable Energy (EERE) through award number DE-EE0009177 provided to the University of Tennessee-Oak Ridge Innovation Institute. This work (S.M.K.) was supported as part of the Institute for Quantum Matter, an Energy Frontier Research Center funded by the U.S. Department of Energy, Office of Science, Basic Energy Sciences under award no. DE-SC0019331. The collection of X-ray data was supported by Prof. Collin Broholm and was performed at the X-ray Crystallography Facility of Johns Hopkins University (JHU) under the help of Maxime A. Siegler. 

\end{acknowledgments}

\section*{Author Contribution} A.B. and Y.W. conceived the project. J.A. developed pc-VAE-BO framework for pyrochlore model. H.D. expanded into pc-VAE-SAASBO framework for pyrochlore model. A.B. supervised J.A and H.D for architecture development. Y.W. collected single-crystal x-ray diffraction data. Y.W. supervised A.M., Y.T., and E.M. to design the pyrochlore simulation model in GSAS-II and generate data from pc-VAE training. S.M.K. grown the single crystal sample of Ho$_2$Ti$_2$O$_7$. B.H.T. supports the development of the GSAS-II software package, including extensions added for this work. A.B. and J.A conducted the analysis, summarized the results and prepared the figures while Y.W. provided continuous feedback on the results. A.B., J.A and Y.W prepared the manuscript. All authors edited the manuscript. 


\section*{Conflict of Interest} The authors confirm there is no conflict of interest.

\section*{Code and Data Availability Statement} The analysis reported here along with the code is summarized in Notebook for the purpose of tutorial and application to other data and can be found in \url{https://github.com/arpanbiswas52/pcVAEBO_pyrochlore}

\section*{References}
\bibliography{aipsamp}

\end{document}